\documentclass[%
reprint,
superscriptaddress,
showpacs,
preprintnumbers,
amsmath,amssymb,
prc,
floatfix]%
{revtex4-1}

\usepackage{color}
\usepackage{CJK}

\usepackage{graphicx}
\usepackage{dcolumn}
\usepackage{bm}
\usepackage[dvipdfmx,bookmarks=true,colorlinks,%
citecolor=blue,linkcolor=blue,anchorcolor=blue,filecolor=blue,urlcolor=blue,%
           ]{hyperref}

\begin{document}


\begin{CJK*}{UTF8}{gbsn}

\title{Systematic study of complete fusion suppression in reactions involving 
weakly bound nuclei at energies above the Coulomb barrier}

\author{Bing Wang (王兵)}%
 \affiliation{Department of Physics, Zhengzhou University, Zhengzhou 
              450001, China}
 \author{Wei-Juan Zhao (赵维娟)}%
 \affiliation{Department of Physics, Zhengzhou University, Zhengzhou 
              450001, China}
\author{Alexis Diaz-Torres}%
 \affiliation{European Centre for Theoretical Studies in Nuclear Physics and 
Related Areas (ECT*), Strada delle Tabarelle 286, I-38123 Villazzano, Trento, 
Italy}
 \author{En-Guang Zhao (赵恩广)}%
 \affiliation{State Key Laboratory of Theoretical Physics,
              Institute of Theoretical Physics, Chinese Academy of Sciences, 
              Beijing 100190, China}
 \affiliation{Center of Theoretical Nuclear Physics, National Laboratory
              of Heavy Ion Accelerator, Lanzhou 730000, China}
\author{Shan-Gui Zhou (周善贵)}%
 \email{sgzhou@itp.ac.cn}
 \affiliation{State Key Laboratory of Theoretical Physics,
              Institute of Theoretical Physics, Chinese Academy of Sciences, 
              Beijing 100190, China}
 \affiliation{Center of Theoretical Nuclear Physics, National Laboratory
              of Heavy Ion Accelerator, Lanzhou 730000, China} 
 \affiliation{Synergetic Innovation Center for Quantum Effects and Application, 
              Hunan Normal University, Changsha, 410081, China}

 \date{\today}

\begin{abstract}
Complete fusion excitation functions of reactions involving breakup are studied 
by using the empirical coupled-channel (ECC) model with breakup effects 
considered. An exponential function with two parameters is adopted to describe 
the prompt-breakup probability in the ECC model. 
These two parameters are fixed by fitting the measured prompt-breakup probability or the 
complete fusion cross sections. The suppression of complete fusion at 
energies above the Coulomb barrier is studied by comparing the data with the 
predictions from the ECC model without the breakup channel considered. The 
results show that the suppression of complete fusion are roughly 
independent of the target for the reactions involving the same projectile. 
\end{abstract} 

\pacs{25.60.Pj, 24.10.-i, 25.70.Mn, 25.70.Jj}

 
\maketitle

\end{CJK*}

\section{\label{sec:introduction}Introduction}

In recent years, the investigation of the breakup effect of weakly bound
nuclei on fusion process 
has been an interesting topic 
\cite{Canto2015_PR596-1,Keeley2007_PPNP59-579,Back2014_RMP86-317}. 
Different processes can take place in collisions involving weakly bound nuclei. 
One is the direct complete fusion (DCF). In this case, the whole projectile 
fuses with the target without breakup. Several processes can occur after the 
breakup of the weakly bound projectile nucleus. When all the fragments fuse with the 
target, the process is called sequential complete fusion (SCF). If only part of 
the fragments fuses with the target nucleus, it is called incomplete fusion 
(ICF). There is also some possibility that none of the fragments is captured by 
the target. This process is called non-capture breakup (NCBU). 

Experimentally, the SCF is difficult to be distinguished from the DCF, as 
the produced compound nuclei from these two processes are the same. Therefore, 
only the complete fusion (CF) cross section, which includes both DCF and SCF 
cross sections, i.e., $\sigma_{\rm CF} = \sigma_{\rm SCF}+\sigma_{\rm DCF}$,  
can be measured. In addition, it is difficult to 
measure separately ICF and CF cross sections owing to the characteristics of 
the evaporation of the excited compound nuclei. For light reaction systems, the 
produced compound nuclei have a large probability for emitting charged 
particles during the cooling process and consequently residues from ICF 
coincide with those from CF. Hence only the sum of the CF and ICF cross 
sections, which is called total fusion (TF) cross section, i.e., 
$\sigma_{\rm TF} \equiv \sigma_{\rm CF} + \sigma_{\rm ICF}$, can be measured. 
For heavy reaction systems, the evaporation of the excited compound nuclei 
occurs mainly by the emission of neutrons and $\alpha$-particles. In this case, 
the separate measurements of CF and ICF cross sections can be 
achieved~\cite{Dasgupta2004_PRC70-024606}. In recent years, many measurements 
of the CF cross sections have been performed 
\cite{Liu2005_EPJA26-73,Gomes2009_NPA828-233,
Shrivastava2009_PRL103-232702,Fang2013_PRC87-024604,Tripathi2005_PRC72-017601, 
Gasques2009_PRC79-034605,Kalita2011_JPG38-095104,
Zhang2014_PRC90-024621,Fang2015_PRC91-014608,Hu2015_PRC91-044619,
Rath2009_PRC79-051601R,Guo2015_PRC92-014615}.

Theoretically, the influence of breakup on the fusion cross section has been an 
extensively studied topic. The continuum-discretized coupled-channels (CDCC) 
framework has been very successful 
\cite{Hagino2000_PRC61-037602,Diaz-Torres2002_PRC65-024606,
Diaz-Torres2003_PRC68-044607,Otomar2013_PRC87-014615,
Beck2007_PRC75-054605,Keeley2001_PRC65-014601}, since it provides a good 
description of observed NCBU, elastic and TF cross sections. However, most CDCC 
calculations have a shortcoming 
\cite{Thompson2004_PTPSupp154-69,Marta2014_PRC89-034625}, as they cannot give the 
ICF and CF cross sections unambiguously \cite{Boselli2014_JPG41-094001}. 
This shortcoming can be avoided within a new dynamical quantum approach that 
includes SCF as well as ICF from the bound state(s) of the projectile 
\cite{Boselli2015_PRC92-044610}. In addition, the comparison of experimental 
fusion cross sections with either the predictions of 
coupled-channel (CC) calculations 
without the breakup and transfer channels 
\cite{Zhang2014_PRC90-024621,Rath2009_PRC79-051601R,
Kumawat2012_PRC86-024607,Fang2015_PRC91-014608,Hu2015_PRC91-044619,
Guo2015_PRC92-014615,Rath2012_NPA874-14,Pradhan2011_PRC83-064606,
Palshetkar2014_PRC89-024607,Rath2013_PRC88-044617} or the predictions of a 
single barrier penetration model (SBPM) \cite{Tripathi2005_PRC72-017601, 
Gasques2009_PRC79-034605,Kalita2011_JPG38-095104} shows
that CF cross sections are suppressed owing to the breakup at energies above 
the Coulomb barrier. 

Many efforts have been made to investigate the systematics of the CF suppression
\cite{Gasques2009_PRC79-034605,Dasgupta2010_NPA834-147c,Wang2014_PRC90-034612,
Rafiei2010_PRC81-024601,Sargsyan2012_PRC86-054610,Gomes2011_PRC84-014615}.
In Ref.~\cite{Rafiei2010_PRC81-024601}, a three-dimensional classical dynamical 
reaction model 
\cite{Diaz-Torres2007_PRL98-152701,Diaz-Torres2010_JPG37-075109,
Diaz-Torres2011_CPC182-1100} together with the measured prompt-breakup 
probabilities was adopted to study the systematics of the suppression for the 
reactions induced by $^{9}$Be. It was found that the CF suppression is nearly 
independent of the target. In Ref.~\cite{Wang2014_PRC90-034612}, a large number 
of CF excitation functions of reactions including the breakup channel were 
studied by applying the universal fusion function prescription 
\cite{Canto2009_JPG36-015109,Canto2009_NPA821-51} with the double folding and 
parameter-free S\~ao Paulo potential 
\cite{CandidoRibeiro1997_PRL78-3270,Chamon1997_PRL79-5218,
Chamon2002_PRC66-014610}. The authors concluded that the CF cross sections are 
suppressed owing to the prompt breakup of projectile and the suppression effect 
for reactions induced by the same projectile is roughly independent of the 
target. 

Recently, a systematic study of capture (fusion) excitation functions for 217 
reaction systems has been performed by using an empirical coupled-channel (ECC) 
model \cite{Wang2015_arXiv1504.00756}. In the ECC model, a barrier distribution 
is used to take effectively into account the effects of couplings to inelastic 
excitations and neutron transfer channels 
\cite{Wang2015_arXiv1504.00756,Wang2016_SciChinaPMA_in-press}. However, the 
coupling to breakup 
channel has not been taken into account. The sub-barrier prompt-breakup 
probabilities for the reactions induced by ${}^{9}$Be have been measured and 
the radial dependence of the breakup probabilities have been 
established~\cite{Hinde2002_PRL89-272701,Rafiei2010_PRC81-024601,
Diaz-Torres2007_PRL98-152701}. In the present work, the characteristics of the 
measured prompt-breakup probability and its effect on CF will be considered in 
the ECC model. In Ref.~\cite{Wang2014_PRC90-034612}, it was found that the 
suppression of CF is sensitive to the lowest breakup threshold energy of the 
projectile and there holds an exponential relation between the suppression and 
the breakup threshold energy. In the present work, a systematic study of the CF 
excitation functions for reactions involving weakly bound nuclei, as well as, of 
the suppression on CF excitation functions will be investigated.

The paper is organized as follows. In Sec.~\ref{sec:methods}, we introduce
the ECC model considering the breakup effect (in short, the ECCBU model). 
In Sec.~\ref{sec:results}, the ECCBU model is applied to analyze the 
data of different projectile induced reactions. The suppression of CF 
excitation functions at energies above the Coulomb barrier will be also 
investigated. Finally, a summary is given in Sec.~\ref{sec:summary}.
 
\section{\label{sec:methods}Method}

The fusion cross section at a given center-of-mass energy $E_{\rm c.m.}$
can be written as the sum of the cross section for each partial wave $J$,
\begin{equation}\label{eq:sig_cap}
\sigma_{\rm Fus}(E_{\rm c.m.})=\pi\lambdabar^2 
                             \sum_{J}^{J_{\rm max}}(2J+1)T(E_{\rm c.m.},J),
\end{equation}
where $\lambdabar^2= \hbar^2/(2\mu E_\mathrm{c.m.})$ is the reduced de Broglie 
wavelength. $\mu$~denotes the reduced mass of the reaction system. $T$ 
denotes the penetration probability 
of the potential barrier between the colliding nuclei at a given $J$. $J_{\rm 
max}$ is the critical angular momentum.

When one of the colliding nuclei is weakly bound, the additional breakup degree 
of freedom makes the colliding process more complicated. A prompt-breakup 
probability $P_{\rm BU}$ is introduced. Considering the survival of projectile 
against breakup before fusion, the CF cross section can be calculated by 
Eq.~(\ref{eq:sig_cap}) with the penetration probability $T$ multiplied by the 
survival probability $(1-P_{\rm BU})$, which is written as 
\cite{Hussein1992_PRC46-377,Hussein1993_PRC47-2398,Canto1995_PRC52-2848R,
Diaz-Torres2002_NPA703-83}
\begin{equation}\label{eq:cf}
 \sigma_{\text{CF}}({E_\mathrm{c.m.}})  \! =  \! \pi\lambdabar^2   \!\! 
    \sum_{J}^{J_{\rm max}}(2J+1)T(E_\mathrm{c.m.},J) 
     [1-P_{\rm BU}(E_\mathrm{c.m.},J)].
\end{equation}
Equation (\ref{eq:cf}) does not include the SCF component which seems 
not to be significant at Coulomb barrier energies 
\cite{Boselli2014_JPG41-094001}. Based on both measurements 
\cite{Hinde2002_PRL89-272701,Rafiei2010_PRC81-024601} and CDCC 
calculations~\cite{Diaz-Torres2007_PRL98-152701}, the breakup probability along 
a given classical orbit can be written as an exponential function of the 
distance of closest approach $R_{\rm min}(E_\mathrm{c.m.},J)$,
\begin{equation}\label{eq:bu}
P_{\rm BU}(E_\mathrm{c.m.},J) = \exp[\nu + \mu R_{\rm min}(E_\mathrm{c.m.},J)],
\end{equation}
where $\nu$ and $\mu$ are the logarithmic intercept and slope parameters of the 
function, respectively. These two parameters can be determined by reproducing 
the measured prompt-breakup probability. 
$R_{\rm min}(E_\mathrm{c.m.},J)=R_{\rm 
B}(J)$ and $R_{\rm B}(J)$ is the position of the 
barrier~\cite{Diaz-Torres2007_PRL98-152701}.

The penetration probability $T$ in Eqs.~(\ref{eq:sig_cap}) and (\ref{eq:cf}) 
is calculated with 
the ECC model in which the coupled-channel effects 
(excluding the breakup channel) are taken into account by introducing a barrier 
distribution $f(B)$ \cite{Wang2015_arXiv1504.00756} 
\begin{equation}\label{eq:Tran}
 T(E_{\rm c.m.},J) =\int f(B)T_{\rm HW}(E_{\rm c.m.},J,B){\rm d}B ,
\end{equation} 
where $B$ is the barrier height. $T_{\rm HW}$ denotes the penetration 
probability calculated by the well-known Hill-Wheeler formula 
\cite{Hill1953_PR089-1102}. 
Note that for very deep sub-barrier penetration, the Hill-Wheeler formula is not
valid because of the long tail of the Coulomb potential. In 
Ref.~\cite{Li2010_IJMPE19-359}, a new barrier penetration formula was
proposed for potential barriers containing a long-range Coulomb interaction
and this formula is especially appropriate for the barrier penetration
with incident energy much lower than the Coulomb barrier. 
The implementation of this barrier penetration formula in the ECC model is in progress. 

The barrier distribution $f(B)$ is taken to be an asymmetric Gaussian function
\begin{equation}\label{eq:distri}
f(B)=\left\{
      \begin{array}{cc}
       \frac1N\exp\left[-\left(\frac{B-B_{\rm m}}{\varDelta_1}\right)^2\right],
                                                  \quad & B < B_{\rm m}, \\[1em]
       \frac1N\exp\left[-\left(\frac{B-B_{\rm m}}{\varDelta_2}\right)^2\right],
                                                  \quad & B > B_{\rm m}.
      \end{array}
     \right. 
\end{equation}
$f(B)$ satisfies the normalization condition $\int f(B)dB=1$. $N 
=\sqrt{\pi}(\varDelta_1+\varDelta_2)/2 $~is a normalization coefficient. 
$\varDelta_1$, $\varDelta_2$, and $B_{\rm m}$ denote the left width, the right 
width, and the central value of the barrier distribution, respectively.

Within the ECC model \cite{Wang2015_arXiv1504.00756}, the barrier distribution 
is related to the couplings to low-lying collective vibrational states, 
rotational states and positive $Q$-value neutron transfer (PQNT) channels. 
The vibrational modes are connected to the change of nuclear shape while the 
nuclear rotational states are related to the static deformations of the 
interacting nuclei. Furthermore, when the two nuclei come close enough to each 
other, both nuclei are distorted owing to the attractive nuclear force and the 
repulsive Coulomb force, thus dynamical deformation develops 
\cite{Wang2012_PRC85-041601R,Zagrebaev2003_PRC67-061601R}.
Considering the dynamical deformation, a two-dimensional potential energy 
surface (PES) with respect to relative distance $R$ and quadrupole deformation 
of the system can be obtained. Based on this PES, empirical formulas for 
calculating the parameters of the barrier distribution were proposed to take 
into account the effect of the couplings to inelastic excitations in 
Ref.~\cite{Wang2015_arXiv1504.00756}.

The effect of the coupling to the PQNT channels is simulated by broadening the 
barrier distribution. Only one neutron pair transfer channel is 
considered in the present model. When the $Q$ value for one neutron pair 
transfer is positive, the widths of the barrier distribution are calculated as 
$ \varDelta_i \rightarrow gQ(2n)+\varDelta_i, (i=1,2)$, where $Q(2n)$ is the 
$Q$ value for one neutron pair transfer. $g$ is taken as $0.32$ for all 
reactions with positive $Q$ value for one neutron pair transfer channel 
~\cite{Wang2015_arXiv1504.00756,Wang2016_SciChinaPMA_in-press}.

\begin{figure}[tb!]
\centering{
\includegraphics[width=0.75\columnwidth]{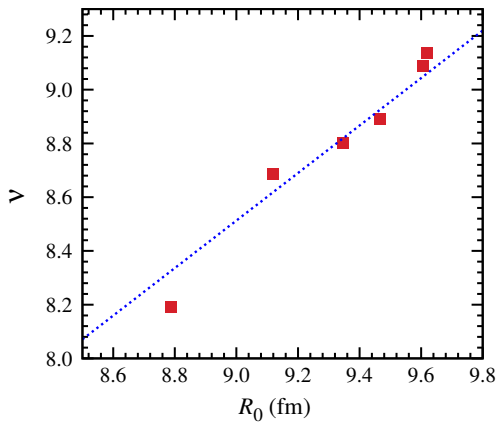} 
 }
\caption{(Color online) The fitted results of $\nu$ shown as a 
function of $R_0$, i.e., $R_0=R_{\rm P}+R_{\rm T}$. The fitted 
results of $\nu$ taken from Ref.~\cite{Rafiei2010_PRC81-024601} are represented 
by the solid squares. The blue dotted line denotes the results obtained from the 
function of $\nu= a - \mu R_0$ with $a=0.557$ and $\mu=-0.884~{\rm 
fm}^{-1}$.}\label{fig:v}
\end{figure}

\section{\label{sec:results}Results and Discussions}

To calculate the complete fusion cross sections for a given reaction, two additional 
parameters $\nu$ and $\mu$ are needed in the ECCBU model. 
These two parameters can be 
determined by reproducing the measured prompt-breakup probability, if 
such measurements are available. 
In Ref.~\cite{Rafiei2010_PRC81-024601}, the sub-barrier prompt-breakup 
probabilities for the reactions induced by ${}^{9}$Be have been measured and the 
radial dependence of the sub-barrier breakup probabilities has been established. 
Therefore, for reactions induced by ${}^{9}$Be, both $\nu$ and $\mu$ can be 
extracted from the measured breakup probabilities. Next, we will first extract 
$\nu$ and $\mu$ from the measured prompt-breakup probabilities. Then these 
$\nu$ and $\mu$ values will be adopted as inputs for the ECCBU calculations.

\begin{figure*}[tb!]
\centering{
\includegraphics[width=0.5\columnwidth]{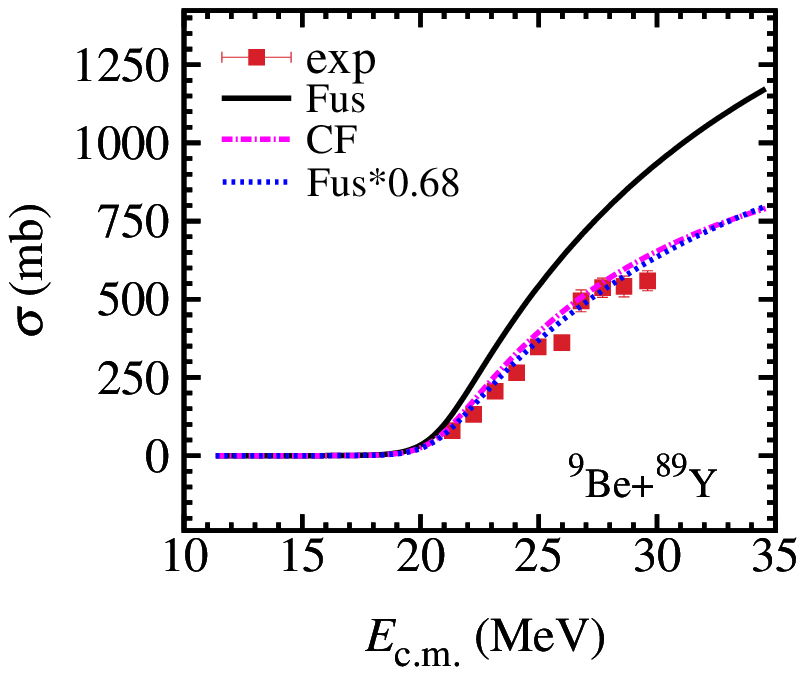}
\includegraphics[width=0.5\columnwidth]{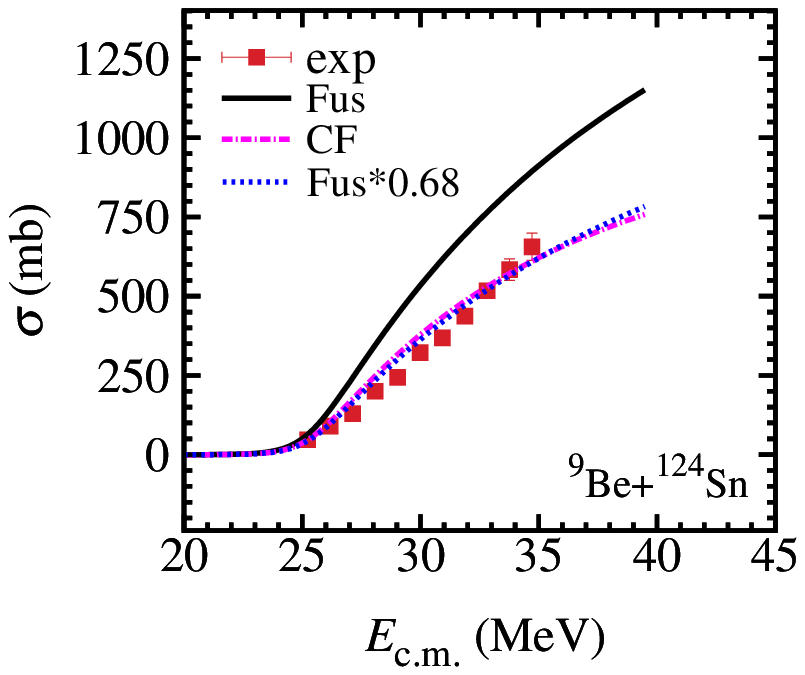}
\includegraphics[width=0.5\columnwidth]{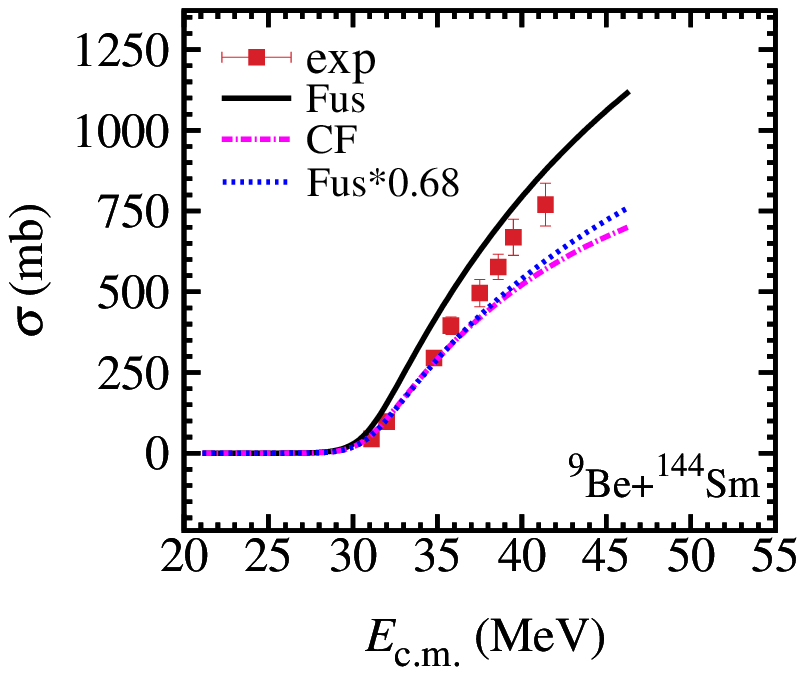}
\includegraphics[width=0.5\columnwidth]{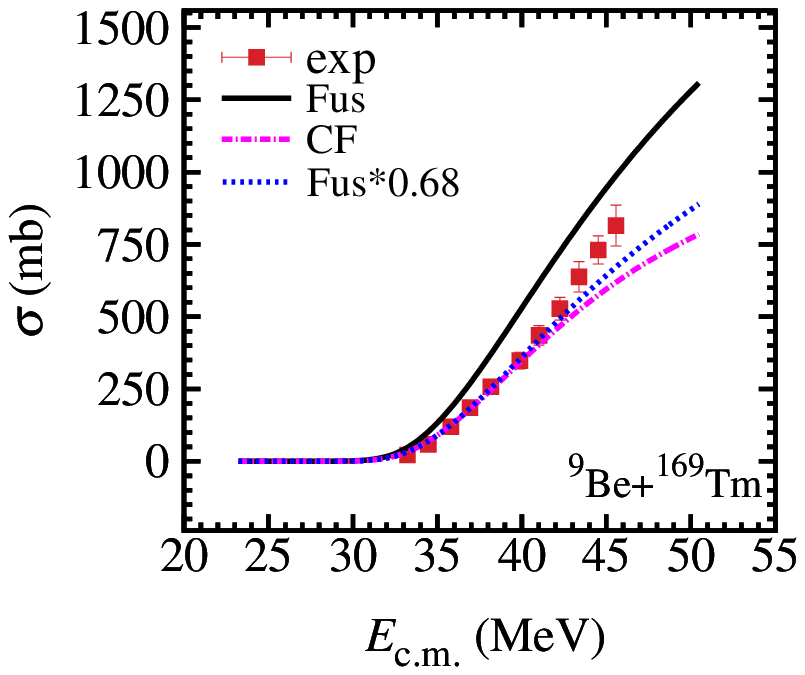}\\
\includegraphics[width=0.5\columnwidth]{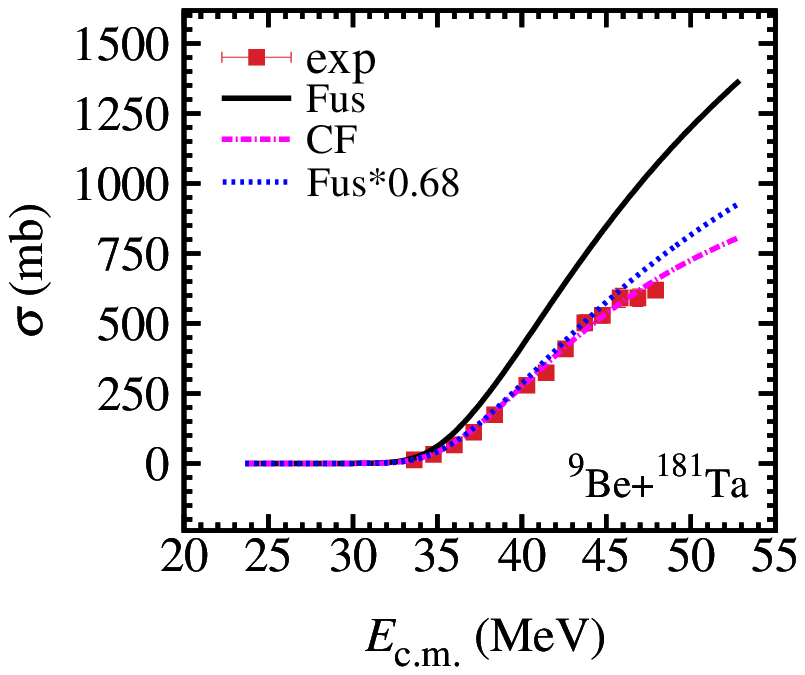}
\includegraphics[width=0.5\columnwidth]{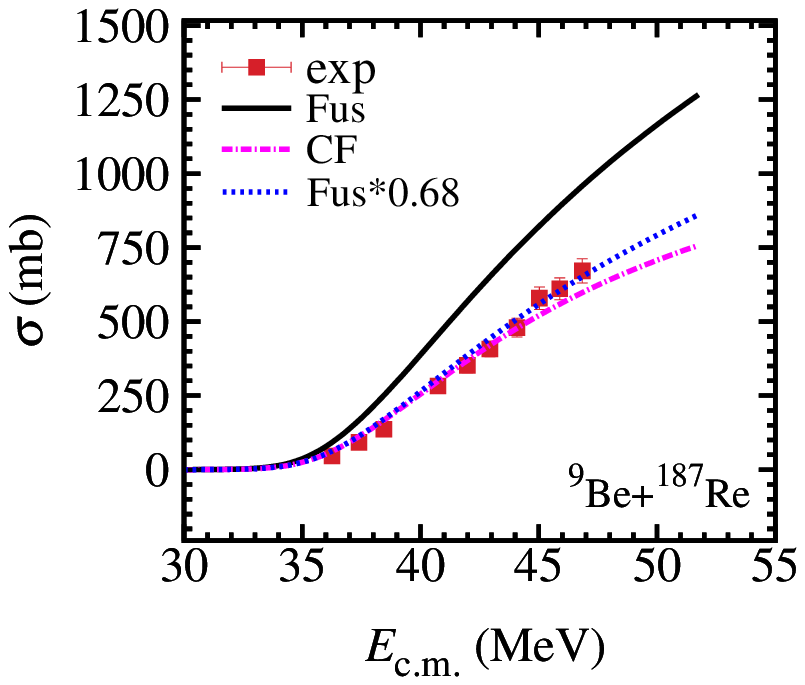}
\includegraphics[width=0.5\columnwidth]{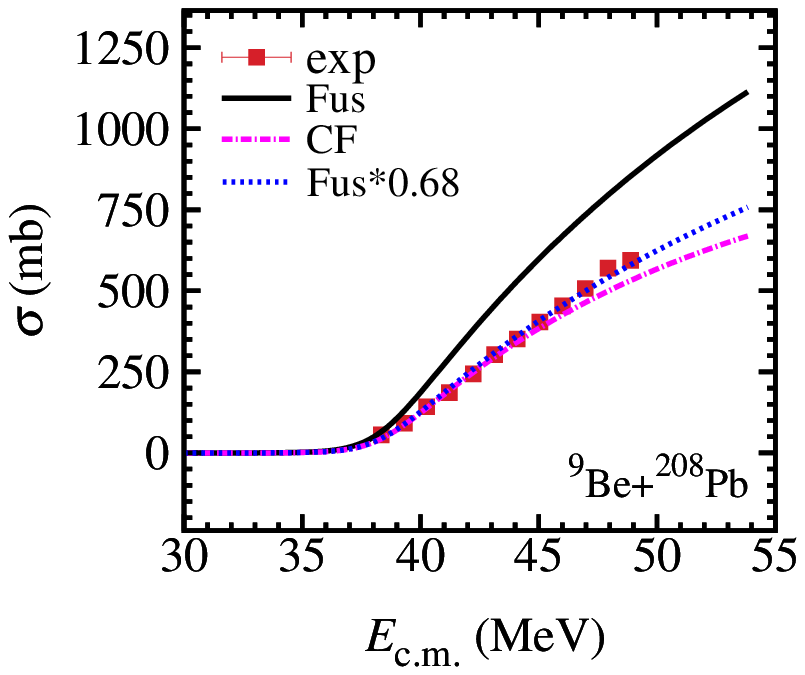}
\includegraphics[width=0.5\columnwidth]{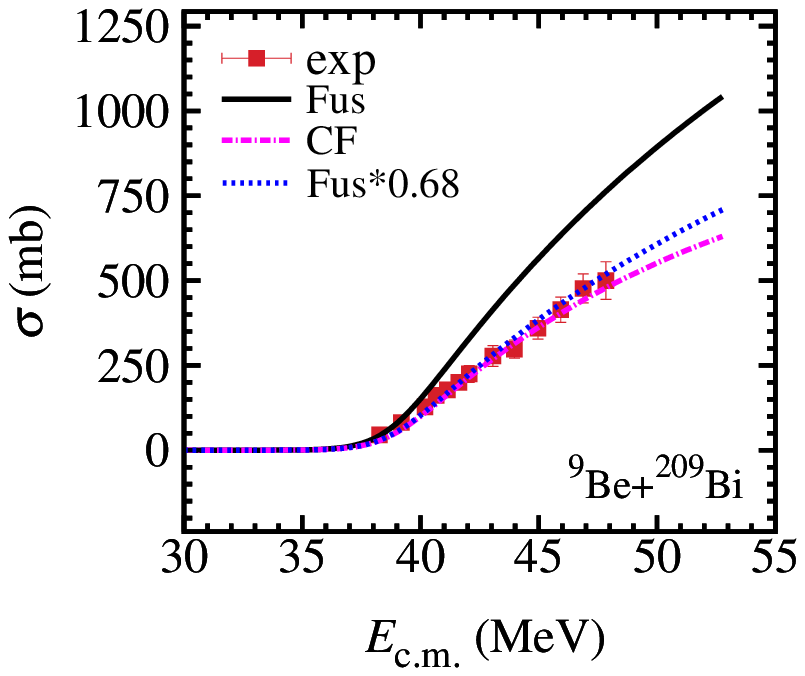}}
\caption{(Color online) The experimental complete fusion 
excitation 
functions and calculated cross sections for reactions induced by ${}^{9}$Be on 
${}^{89}$Y~\cite{Palshetkar2010_PRC82-044608}, 
${}^{124}$Sn~\cite{Parkar2010_PRC82-054601}, 
 ${}^{144}$Sm~\cite{Gomes2006_PRC73-064606,Gomes2009_NPA828-233},
 ${}^{169}$Tm~\cite{Fang2015_PRC91-014608}, 
 ${}^{181}$Ta~\cite{Zhang2014_PRC90-024621},
 ${}^{187}$Re~\cite{Fang2015_PRC91-014608}, 
 ${}^{208}$Pb~\cite{Dasgupta2004_PRC70-024606}, and 
${}^{209}$Bi~\cite{Signorini1998_EPJA2-227,Dasgupta2010_PRC81-024608}. 
The black line (Fus) denotes the fusion cross sections obtained from 
the ECC model without the breakup channel considered, i.e., $\sigma_\mathrm{Fus}$. 
The pink dash-dotted line 
(CF) denotes the calculated complete fusion cross sections obtained from 
the ECCBU model with $\nu= 0.557-\mu R_0$ and $\mu =-0.884$~fm$^{-1}$. 
The blue dotted line denotes $F_\mathrm{B.U.} \sigma_\mathrm{Fus}$ with the suppression 
factor $F_\mathrm{B.U.}=0.68$ taken from Ref.~\cite{Wang2014_PRC90-034612}.}
\label{fig:9Be}
\end{figure*}

\subsection{\label{subsec:com}Complete fusion for reactions involving 
the weakly bound projectile ${}^{9}$Be}

In Ref. \cite{Rafiei2010_PRC81-024601}, it was found that when the breakup 
probabilities are presented as a function of the surface separation of the two 
interacting nuclei, i.e., $R_{\rm min}- R_0$, the effect of nuclear 
size is removed and the breakup probabilities for all targets 
with $62 < Z < 83$ overlap. Here $R_0=R_{\rm P}+R_{\rm T}$ is the summed 
radius of the interacting nuclei. $R_{\rm P}$ and $R_{\rm T}$ are the radii of 
the equivalent spherical nuclei and calculated using $R_{\rm P(T)} = r_0 
A^{1/3}_{\rm P(T)}$ with $r_0=1.2$ fm \cite{Rafiei2010_PRC81-024601}. 
This implies that the prompt-breakup probability can be written as
\begin{equation}\label{eq:bu1}
P_{\rm BU} = \exp[\nu + \mu R_0 + \mu (R_{\rm 
min}-R_0)],
\end{equation}
with both the $\nu + \mu R_0$ and slope $\mu$ are independent of the 
target. Therefore, $\nu$ should satisfy $\nu= a - \mu R_0$. Then 
Eq.~(\ref{eq:bu1}) becomes 
\begin{equation}\label{eq:bu2}
P_{\rm BU} = \exp[a + \mu (R_{\rm min}-R_0)].
\end{equation}
$\mu$ and $a$ are independent of the target and can be extracted by making a fit 
to the measured prompt-breakup probabilities. Actually, in 
Ref.~\cite{Rafiei2010_PRC81-024601}, the target-independent slope parameter 
$\mu$ was given as $\bar\mu=-0.884\pm0.011$~fm$^{-1}$. Meanwhile, the values 
of $\nu$ for the reactions were obtained by fitting the measured prompt-breakup 
probabilities using the mean slope $\mu=-0.884$~fm$^{-1}$, which are shown 
as a function of $R_0$ in Fig.~\ref{fig:v}. As discussed above, $\nu$ should 
satisfy the function of $\nu= a - \mu R_0$. So $a$ can be determined by making a 
fit to the values of $\nu$, and $a=0.557$ is obtained. The results of  
$\nu=0.557-\mu R_0$ are displayed as the blue dotted line in Fig.~\ref{fig:v}.

In the present work, the sub-barrier prompt-breakup probability denoted by 
Eq.~(\ref{eq:bu}) with $\nu=0.557-\mu R_0$ and $\mu=-0.884$~fm$^{-1}$ is 
adopted to perform the ECCBU calculations for the reactions induced by 
${}^{9}$Be.
The reactions with 
${}^{89}$Y~\cite{Palshetkar2010_PRC82-044608}, 
${}^{124}$Sn~\cite{Parkar2010_PRC82-054601}, 
${}^{144}$Sm~\cite{Gomes2006_PRC73-064606,Gomes2009_NPA828-233},
${}^{169}$Tm~\cite{Fang2015_PRC91-014608}, 
${}^{181}$Ta~\cite{Zhang2014_PRC90-024621},
${}^{187}$Re~\cite{Fang2015_PRC91-014608}, 
${}^{208}$Pb~\cite{Dasgupta2004_PRC70-024606}, and 
${}^{209}$Bi~\cite{Signorini1998_EPJA2-227,Dasgupta2010_PRC81-024608} as 
targets have been investigated. 
The comparison of the calculated CF cross sections with the experimental values 
are shown in Fig.~\ref{fig:9Be}. 
The pink dash-dotted line (CF) denotes the calculated CF cross sections. Note 
that after the breakup of $^9$Be into $\alpha+\alpha+n$, 
the capture of either of 2$\alpha$ particles by the target or the SCF can 
also occur and, therefore, contribute to experimental CF cross sections. In the 
present work, we do not take into account these events because this is a very 
complex problem that needs to be further investigated. Hence, the present 
calculations provide a lower limit for CF cross sections. Comparing this lower 
limit with the experimental CF cross sections, one can find that the calculated 
CF cross sections are in good agreement with the data, except the reaction 
${}^{9}$Be + $^{144}$Sm. 
Therefore one can conclude that the capture of all individual 
components of ${}^{9}$Be by the 
target, after the ${}^{9}$Be breakup, is not 
very significant. For convenience, we label the calculated cross sections 
from the ECC model without the breakup channel considered by the subscript 
``Fus'', i.e., $\sigma_{\rm Fus}$.

As mentioned above, a large number of CF excitation functions of reactions 
including the breakup channel have been studied by applying the UFF 
prescription 
\cite{Canto2009_JPG36-015109,Canto2009_NPA821-51} in 
Ref.~\cite{Wang2014_PRC90-034612}. It was found that the suppression 
effect for reactions induced by the same projectile is independent of the 
target. For the reactions involving ${}^{9}$Be, the suppression factor $F_\mathrm{B.U.}$, 
which is defined as the ratio of the data to the UFF, i.e., the predictions from the 
Wong's formula \cite{Wong1973_PRL31-766}, is 0.68, and the reaction ${}^{9}$Be + 
$^{144}$Sm was also not following the systematics found in 
Ref.~\cite{Wang2014_PRC90-034612}. 

As the inelastic excitation couplings are 
not important at energies well-above the Coulomb barrier, the suppression 
obtained from ECC calculations without the breakup channel considered should be 
similar to those obtained from the UFF \cite{Wang2014_PRC90-034612}. 
To check this,  
the predictions from the ECC model without the breakup channel 
considered, which are shown in Fig.~\ref{fig:9Be} by the black line, are used 
as a reference to be compared with the data. One 
can find that the CF cross sections are suppressed as compared with 
$\sigma_{\rm Fus}$ at above-barrier energies.
We scale $\sigma_{\rm Fus}$ by the suppression factor $F_\mathrm{B.U.}=0.68$ and 
show it in Fig.~\ref{fig:9Be} by the blue dotted line. It can be seen that the 
blue dotted line roughly coincides with the data and the pink dash-dotted line, 
while small deviations from the pink dash-dotted line at high energies. 
Therefore, comparing $\sigma_{\rm Fus}$ with the CF cross sections, the 
suppression of CF cross sections for reactions induced by ${}^{9}$Be is 
independent of the target and the suppression factor is about 0.68, which is 
consistent with the result obtained in 
Ref.~\cite{Wang2014_PRC90-034612}.

\begin{figure*}[tb!]
\centering{
\includegraphics[width=0.5\columnwidth]{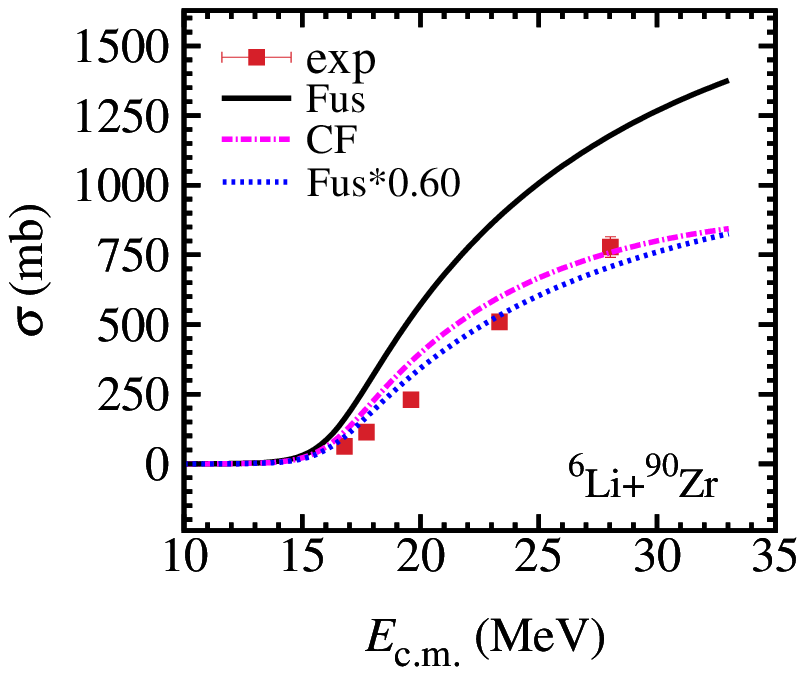}
\includegraphics[width=0.5\columnwidth]{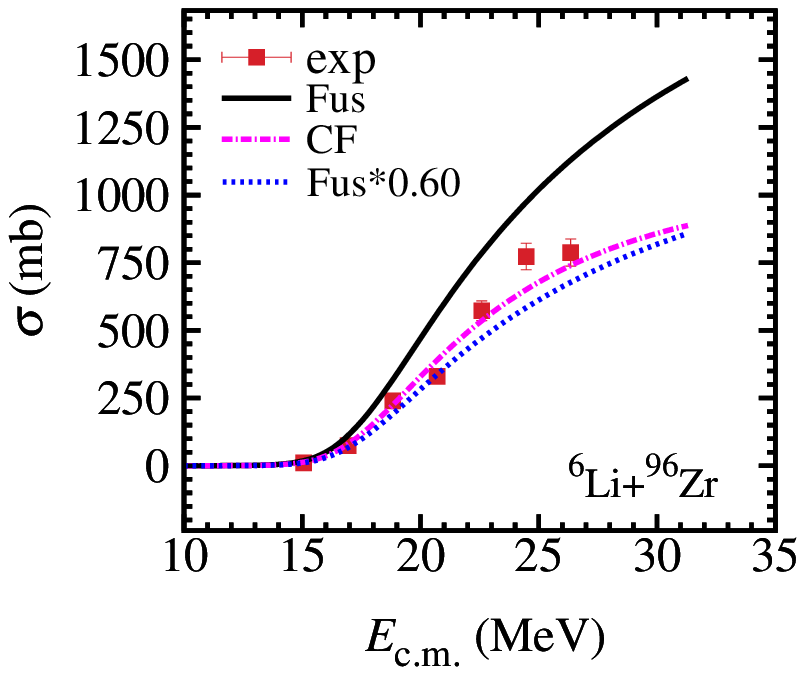}
\includegraphics[width=0.5\columnwidth]{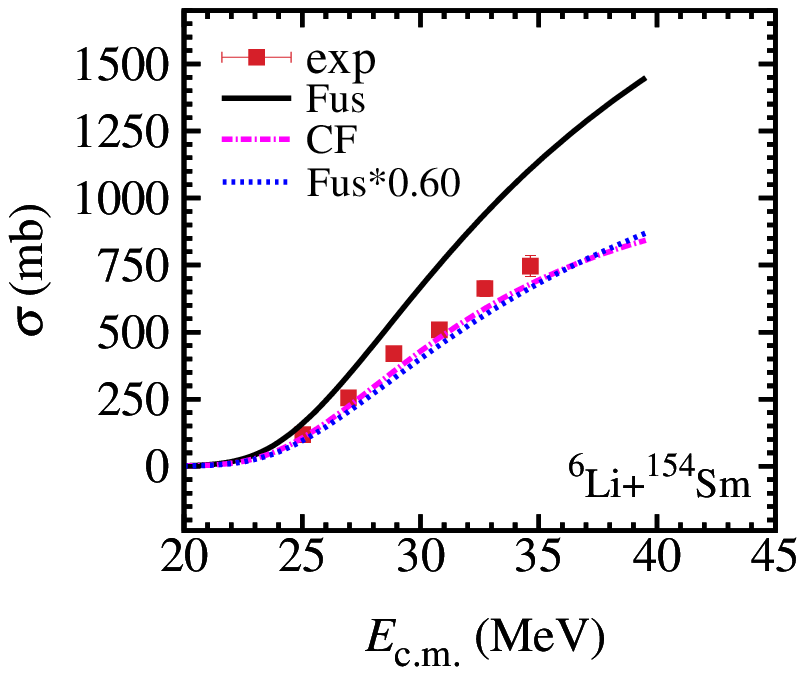}
\includegraphics[width=0.5\columnwidth]{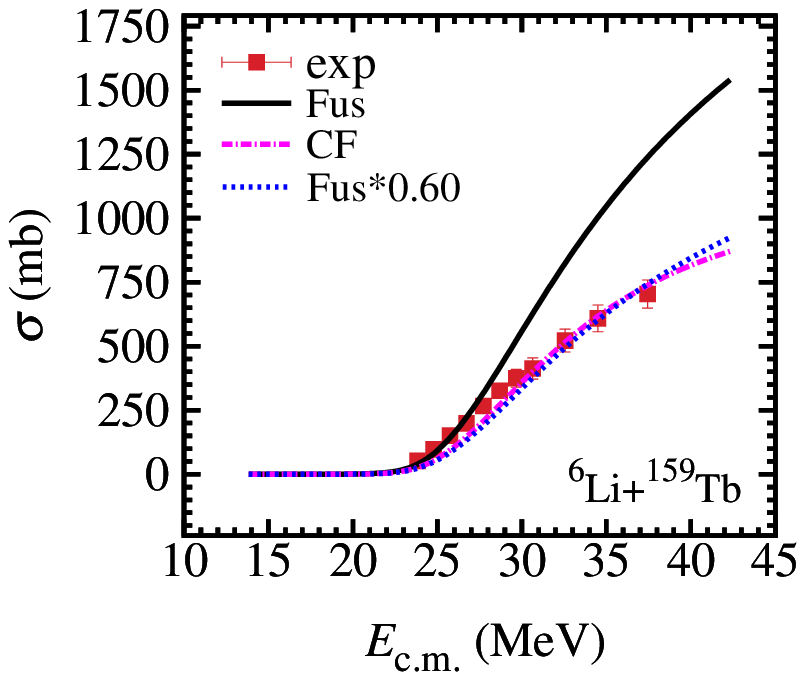}\\
\includegraphics[width=0.5\columnwidth]{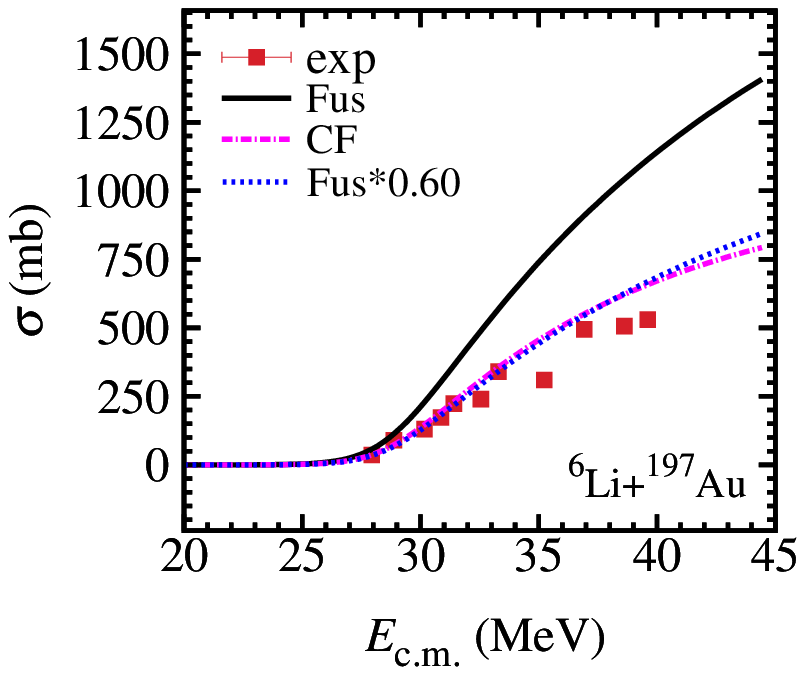}
\includegraphics[width=0.5\columnwidth]{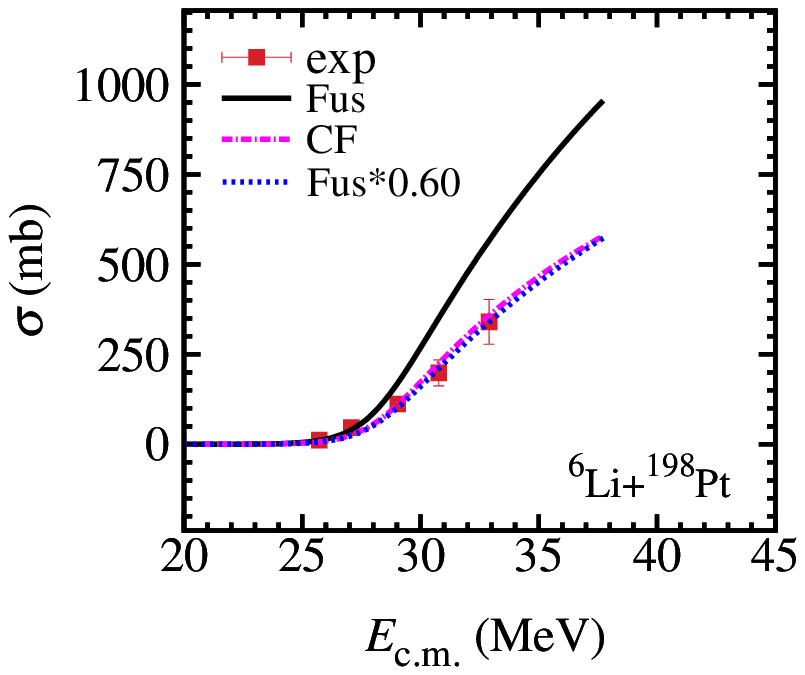}
\includegraphics[width=0.5\columnwidth]{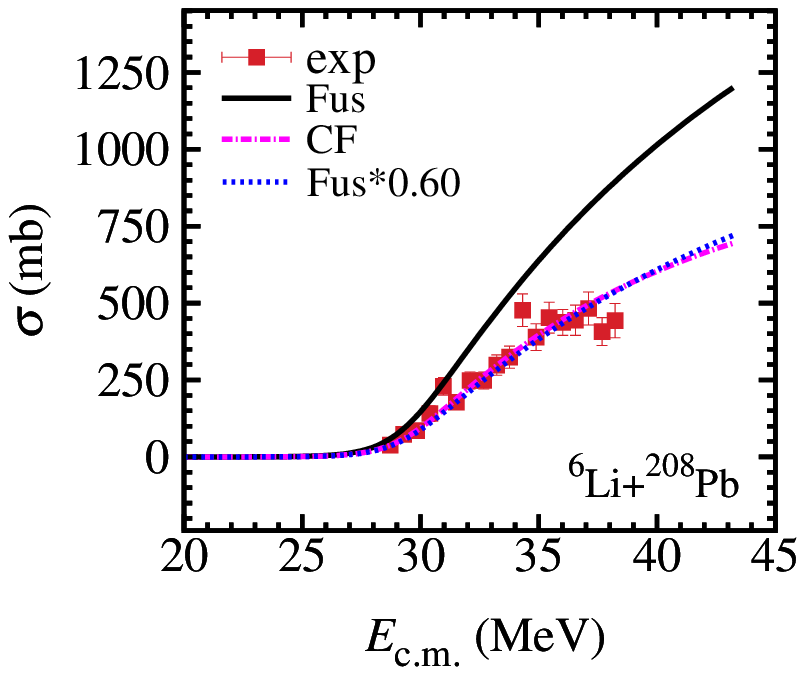}
\includegraphics[width=0.5\columnwidth]{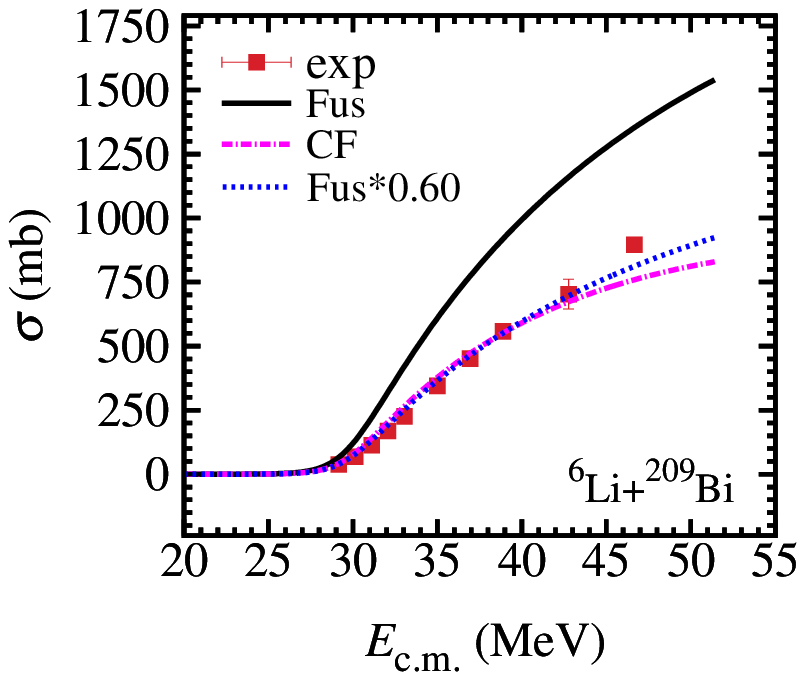}}
\caption{(Color online) The experimental complete fusion 
excitation functions and calculated cross sections for reactions induced by 
${}^{6}$Li on 
${}^{90}$Zr~\cite{Kumawat2012_PRC86-024607},  
${}^{96}$Zr~\cite{Hu2015_PRC91-044619}, 
${}^{154}$Sm~\cite{Guo2015_PRC92-014615},
$^{159}$Tb~\cite{Pradhan2011_PRC83-064606},
$^{197}$Au~\cite{Palshetkar2014_PRC89-024607}, 
$^{198}$Pt~\cite{Shrivastava2009_PRL103-232702},
$^{208}$Pb~\cite{Liu2005_EPJA26-73}, and
$^{209}$Bi~\cite{Dasgupta2004_PRC70-024606}. 
The black line (Fus) denotes the fusion cross sections obtained from 
the ECC model without the breakup channel considered, i.e., $\sigma_\mathrm{Fus}$. 
The pink dash-dotted line 
(CF) denotes the calculated complete fusion cross sections obtained from the ECCBU model 
with $\nu= 0.557-\mu R_0$ and $\mu =-0.798$~fm$^{-1}$. 
The blue dotted line denotes $F_\mathrm{B.U.} \sigma_\mathrm{Fus}$ with the suppression 
factor $F_\mathrm{B.U.}=0.60$ taken from Ref.~\cite{Wang2014_PRC90-034612}.}
\label{fig:6Li}
\end{figure*}
 
These results are very interesting and somehow unexpected, since it is widely 
accepted that the Coulomb breakup should increase with the charge of the target. 
Actually, recent CDCC calculations performed by D. R. Otomar {\it et al.} 
\cite{Otomar2013_PRC87-014615} showed that the total breakup, including the 
interference between its Coulomb and nuclear components, increases with the 
target charge and mass, which seems to be contradictory with the conclusion 
that CF suppression is independent of the target. A possible explanation 
for this apparent contradiction was recently given in 
Refs.~\cite{Wang2014_PRC90-034612,Zhang2014_PRC90-024621,Fang2015_PRC91-014608, 
Hu2015_PRC91-044619}. The breakup may be of two kinds: prompt and delayed, the 
former taking place when the projectile is approaching the target and the 
latter taking place following direct transfer of nucleons or the excitation of 
the projectile to a long-lived resonance above the breakup threshold. The 
experimental results show that the time scale of the delayed breakup is 
several orders of magnitude longer than the collision time and consequently 
only the prompt breakup may affect the fusion processes
\cite{Hinde2002_PRL89-272701,Luong2011_PLB695-105,Luong2013_PRC88-034609}.
In the CDCC calculations made by D. R. Otomar {\it et al.} 
\cite{Otomar2013_PRC87-014615}, both the prompt and delayed breakups
were included. 
In the present work, the measured prompt-breakup 
probabilities are used in the ECCBU calculations and the calculated CF 
cross sections are in good agreement with data. 
Such good agreement supports to some extent this explanation. 
It is highly desirable that one can deal with the prompt and 
delayed breakups separately in CDCC calculations. 
Such calculations will provide further insight in the understanding of 
the target-independent suppression of CF and also provide the prompt-breakup 
probabilities as inputs for the ECCBU calculations.

\begin{table*}[htb!] 
\caption{ %
Slope parameter $\mu$ obtained by making a least-squares fit to the
corresponding CF data using $\nu= 0.557-\mu R_0$ for reactions with 
${}^{6}$Li as projectile.
}
\begin{ruledtabular}
\begin{tabular}{ccccccccr}
 & ${}^{90}$Zr  & ${}^{96}$Zr  &  ${}^{154}$Sm &  ${}^{159}$Tb &  
   ${}^{197}$Au & ${}^{198}$Pt &  ${}^{208}$Pb &  ${}^{209}$Bi \\  
 \hline\noalign{\vskip3pt} 
$\mu$ (fm$^{-1}$) & $-$0.732 & $-$0.803   & $-$0.915  & $-$0.810 & $-$0.711 &
$-$0.810 & $-$0.794 & $-$0.806 \\ 
\end{tabular} \label{table:6Li}
\end{ruledtabular}
\end{table*}

\subsection{\label{subsec:Li}Complete fusion for reactions involving ${}^{6,7}$Li and 
${}^{10,11}$B}

Based on the measured prompt-breakup probabilities for the reactions involving 
${}^{9}$Be, the function for $\nu$ is determined as $\nu=a-\mu R_0$. Meanwhile, 
the suppression of CF cross section at energies above the Coulomb barrier and 
$\mu$ are independent of the target. Next, we will use the ECCBU model with 
$\nu=a-\mu R_0$ to study the reactions induced by ${}^{6}$Li, ${}^{7}$Li, 
${}^{10}$B, and ${}^{11}$B. We assume that $a=0.557$ and SCF is not very 
significant as compared to DCF. For each reaction, the slope parameter $\mu$ 
will be obtained 
by making a fit to the data. Then the systematic behavior of the prompt-breakup 
probabilities and the suppression of CF cross sections will be explored.

\begin{figure}[htb!]
\centering{
\includegraphics[width=0.49\columnwidth]{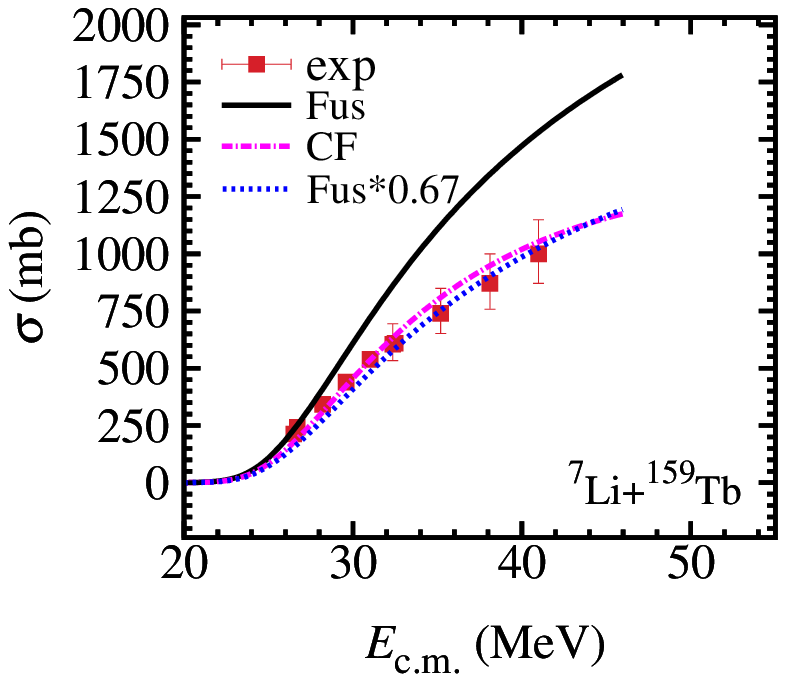}
\includegraphics[width=0.49\columnwidth]{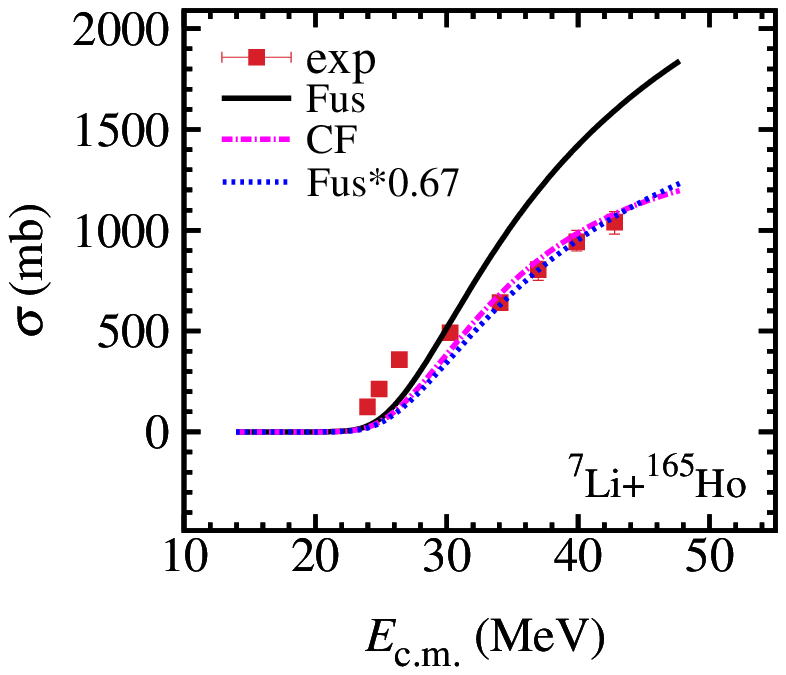}
\includegraphics[width=0.49\columnwidth]{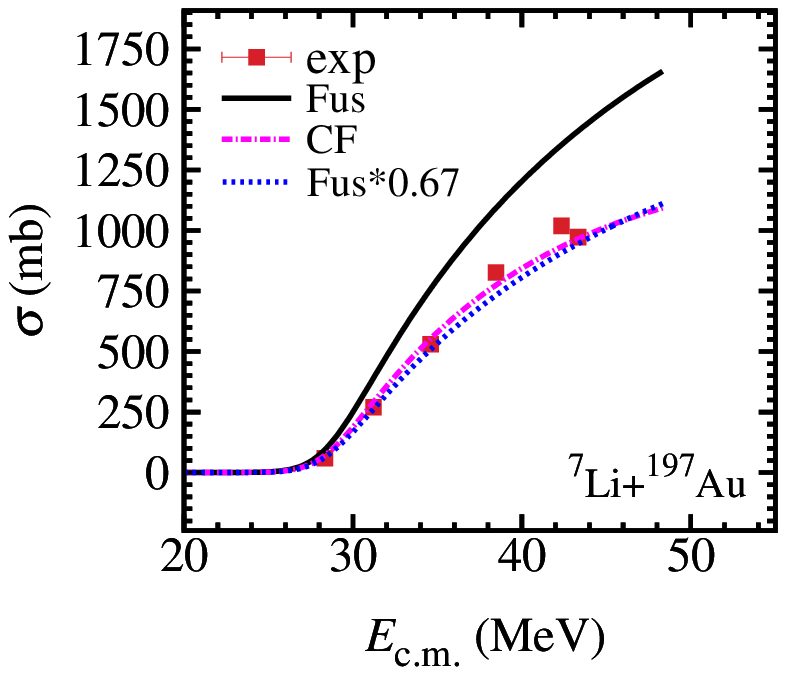}
\includegraphics[width=0.49\columnwidth]{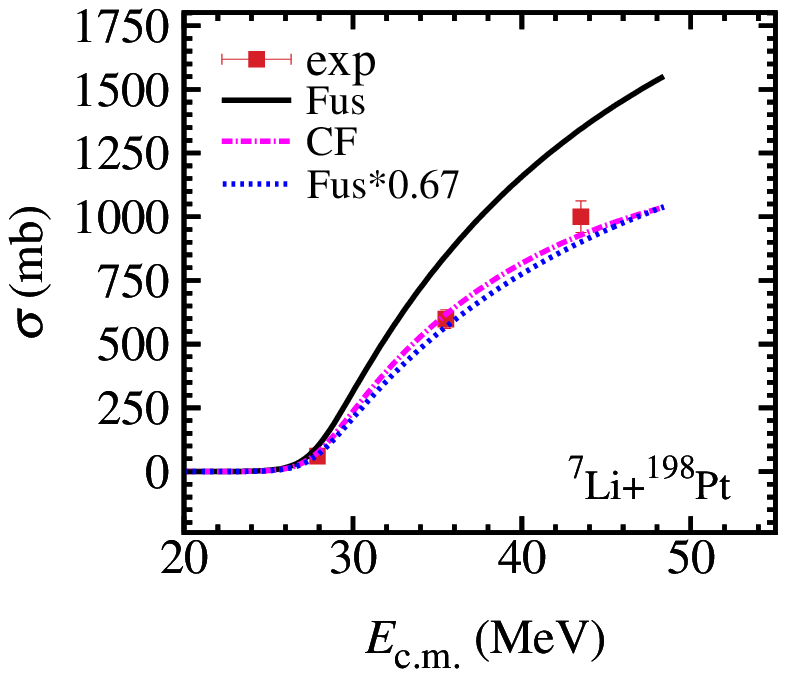}
\includegraphics[width=0.49\columnwidth]{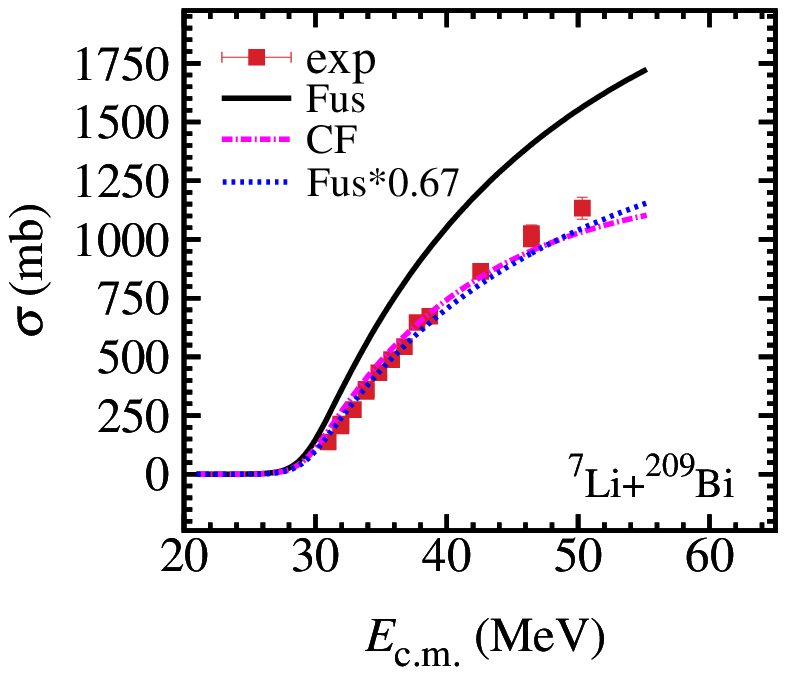}
\phantom{\includegraphics[width=0.48\columnwidth]{7Li209Bi.eps}}
}
\caption{(Color online)The experimental complete fusion 
excitation 
functions and calculated cross sections for reactions induced by ${}^{7}$Li on 
$^{159}$Tb~\cite{Broda1975_NPA248-356}, 
$^{165}$Ho~\cite{Tripathi2002_PRL88-172701,Tripathi2005_PRC72-017601},
$^{197}$Au~\cite{Palshetkar2014_PRC89-024607},
$^{198}$Pt~\cite{Shrivastava2013_PLB718-931}, and 
$^{209}$Bi~\cite{Dasgupta2004_PRC70-024606}. 
The black line (Fus) denotes the fusion cross sections obtained from 
the ECC model without the breakup channel considered, i.e., $\sigma_\mathrm{Fus}$. 
The pink dash-dotted line 
(CF) denotes the calculated complete fusion cross sections obtained from the ECCBU model 
with $\nu= 0.557-\mu R_0$ and $\mu =-0.964$~fm$^{-1}$. 
The blue dotted line denotes $F_\mathrm{B.U.} \sigma_\mathrm{Fus}$ with the suppression 
factor $F_\mathrm{B.U.}=0.67$ taken from Ref.~\cite{Wang2014_PRC90-034612}.}
\label{fig:7Li}
\end{figure}

First the complete fusion excitation functions for reactions induced by 
${}^{6}$Li on 
${}^{90}$Zr~\cite{Kumawat2012_PRC86-024607}, 
${}^{96}$Zr~\cite{Hu2015_PRC91-044619}, 
${}^{154}$Sm~\cite{Guo2015_PRC92-014615},
$^{159}$Tb~\cite{Pradhan2011_PRC83-064606},
$^{197}$Au~\cite{Palshetkar2014_PRC89-024607}, 
$^{198}$Pt~\cite{Shrivastava2009_PRL103-232702},
$^{208}$Pb~\cite{Liu2005_EPJA26-73}, and
$^{209}$Bi~\cite{Dasgupta2004_PRC70-024606} are investigated.
The fitted values of $\mu$ are listed in Table~\ref{table:6Li}.
Similar to the results for ${}^{9}$Be, one can find that the slope parameters 
$\mu$ are also roughly independent of the target, with a mean value of 
$\bar\mu=-0.798$~fm$^{-1}$. Then the mean value of $\mu=-0.798$~fm$^{-1}$ and 
$\nu=0.557-\mu R_0$ are adopted to perform the ECCBU calculations. 
The comparison of the calculated CF cross sections to the experimental values 
are shown in Fig.~\ref{fig:6Li}. The calculated CF cross sections are shown by 
the pink dash-dotted line. It can be seen that the calculated CF cross sections 
are in good agreement with the data. To explore the suppression of CF cross 
sections, the theoretical predictions without the breakup channel considered, 
i.e., $\sigma_{\rm Fus}$ are shown by the black line. 
$\sigma_{\rm Fus}$ multiplied by the suppression factor $F_\mathrm{B.U.}=0.60$ taken from 
Ref.~\cite{Wang2014_PRC90-034612} are represented by the blue dotted line. One 
can find that the results denoted by the blue dotted line are in good agreement 
with the data. It implies that the suppression effect owing to the breakup of 
${}^{6}$Li is independent of the target and the suppression factor should be also 
about 0.60. 

\begin{table}[htb!] 
\caption{ %
Slope parameter $\mu$ obtained by making a least-squares fit to the
corresponding CF data using $\nu= 0.557-\mu R_0$ for reactions with 
${}^{7}$Li as projectile.
}
\begin{ruledtabular}
\begin{tabular}{cccccr}
  & $^{159}$Tb &  ${}^{165}$Ho & ${}^{197}$Au & ${}^{198}$Pt  &  ${}^{209}$Bi \\
 \hline\noalign{\vskip3pt} 
$\mu$ (fm$^{-1}$) & $-$0.956 & $-$0.941 & $-$0.971 & $-$0.993 & $-$0.957  
\\ 
\end{tabular} \label{table:7Li}
\end{ruledtabular}
\end{table}

For ${}^{7}$Li, the experimental complete fusion 
excitation functions for the reactions with 
$^{159}$Tb~\cite{Broda1975_NPA248-356}, 
$^{165}$Ho~\cite{Tripathi2002_PRL88-172701,Tripathi2005_PRC72-017601},
$^{197}$Au~\cite{Palshetkar2014_PRC89-024607},
$^{198}$Pt~\cite{Shrivastava2013_PLB718-931}, and 
$^{209}$Bi~\cite{Dasgupta2004_PRC70-024606} as targets have been measured.
The fitted values of $\mu$ are listed in Table~\ref{table:7Li}. 
Similar to the results for ${}^{9}$Be and ${}^{6}$Li, one can find that the 
slope parameter $\mu$ is also roughly independent of the target, with a 
mean value of $\bar\mu=-0.964$~fm$^{-1}$. Again the mean value of 
$\mu=-0.964$~fm$^{-1}$ and $\nu=0.557-\mu R_0$ are adopted to perform the 
ECCBU calculations. The calculated CF cross sections are shown by 
the pink dash-dotted line in Fig.~\ref{fig:7Li} and are also in good agreement 
with the data. $\sigma_{\rm Fus}$ are shown by the black line. 
We scale $\sigma_{\rm Fus}$ by the suppression factor $F_\mathrm{B.U.}=0.67$ taken from 
Ref.~\cite{Wang2014_PRC90-034612} and show it in Fig.~\ref{fig:7Li} by the blue 
dotted line. One can find that the results denoted 
by the blue dotted line coincide with the data. It implies that the suppression 
of CF cross sections owing to the breakup of ${}^{7}$Li is also independent of 
the target and the suppression factor is about 0.67. Furthermore, the 
suppression for ${}^{7}$Li is weaker than that for ${}^{6}$Li.

\begin{figure}
\centering{
\includegraphics[width=0.49\columnwidth]{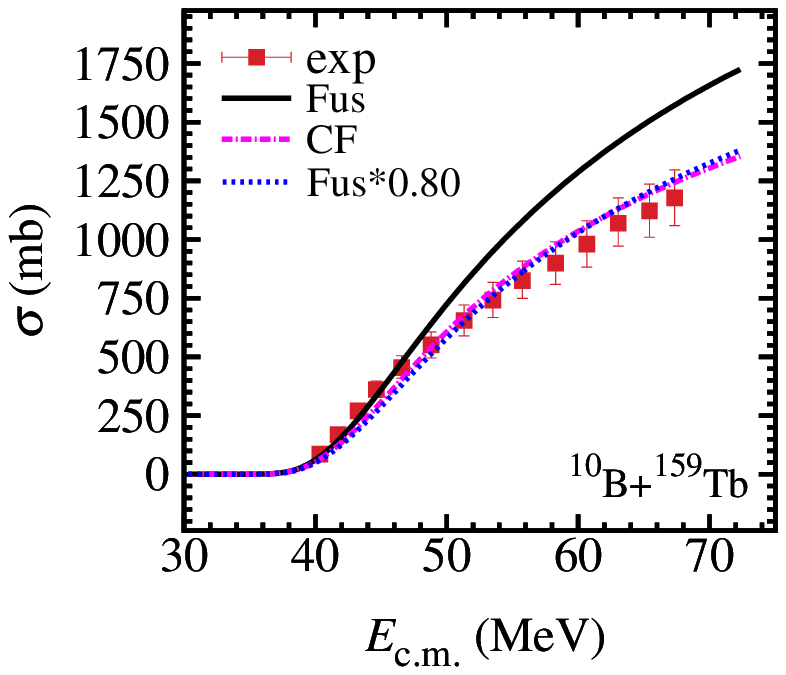}
\includegraphics[width=0.49\columnwidth]{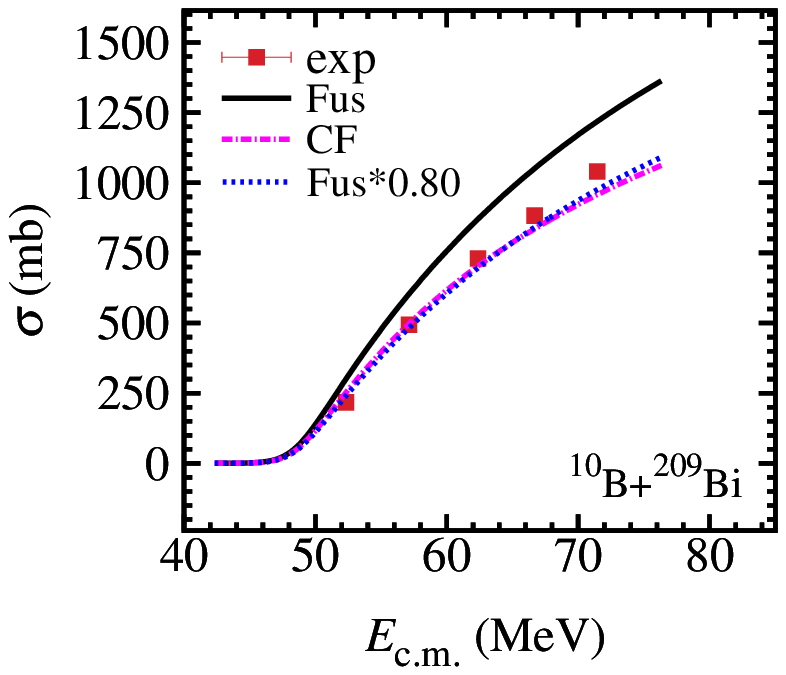}
\includegraphics[width=0.49\columnwidth]{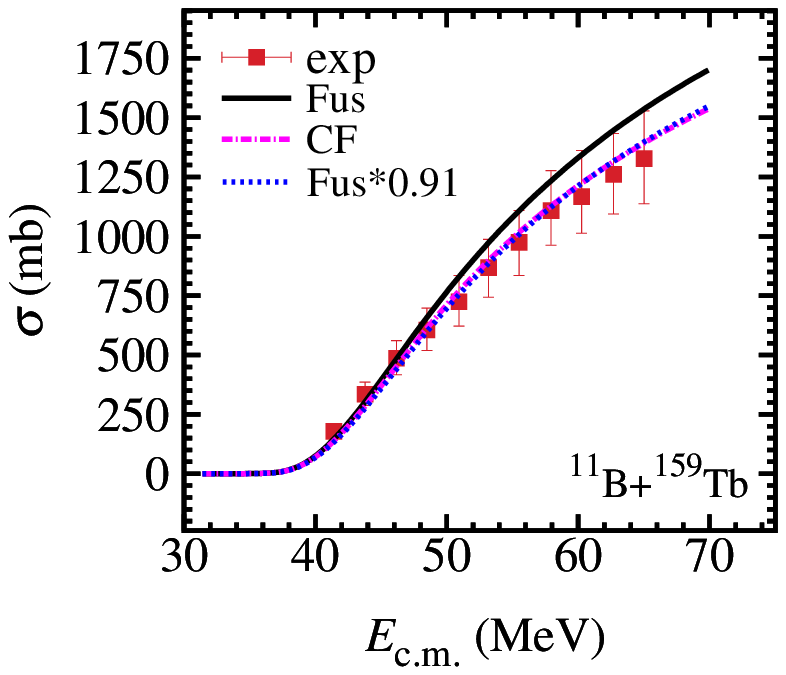}
\includegraphics[width=0.49\columnwidth]{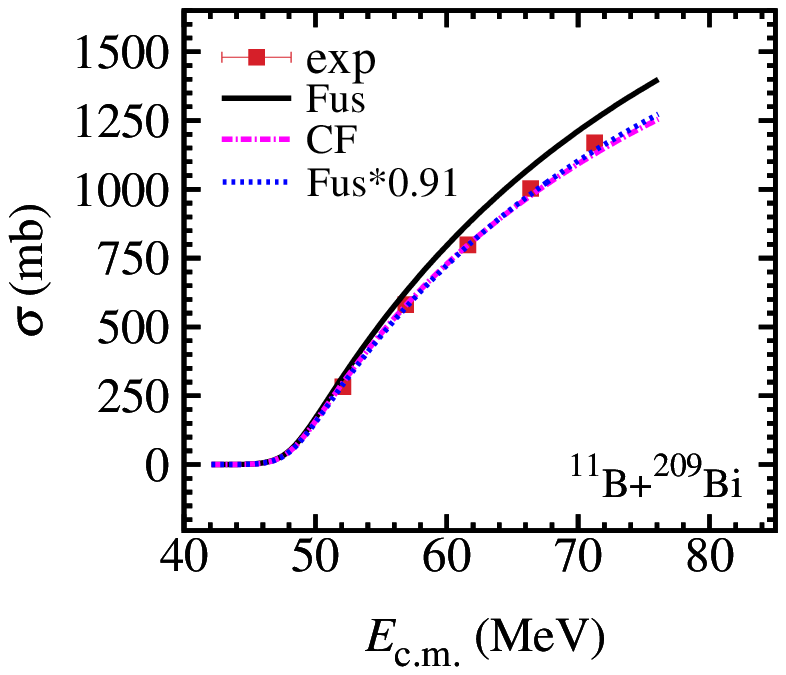}
 }
\caption{(Color online) The experimental complete fusion excitation 
functions and calculated cross sections for reactions induced by ${}^{10,11}$B 
on $^{159}$Tb~\cite{Mukherjee2006_PLB636-91} and 
$^{209}$Bi~\cite{Gasques2009_PRC79-034605}. 
The black line (Fus) denotes the fusion cross section obtained from 
the ECC model without the breakup channel considered, i.e., $\sigma_\mathrm{Fus}$. 
The pink dash-dotted line (CF) denotes the calculated complete 
fusion cross section using the ECCBU model with $\nu= 0.557-\mu R_0$ and $\mu = 
-1.415$~fm$^{-1}$ for ${}^{10}$B, while for ${}^{11}$B, $\mu = 
-1.900$~fm$^{-1}$. The blue dotted line denotes $F_\mathrm{B.U.} \sigma_\mathrm{Fus}$
with 
the suppression factor $F_\mathrm{B.U.}=0.80$ for ${}^{10}$B and 
$F_\mathrm{B.U.}=0.91$ for ${}^{11}$B taken from 
Ref.~\cite{Wang2014_PRC90-034612}.}
\label{fig:B}
\end{figure}

For the reactions involving ${}^{10}$B and ${}^{11}$B, the fitted 
values of $\mu$ for ${}^{10,11}$B + ${}^{159}$Tb~\cite{Mukherjee2006_PLB636-91} 
and ${}^{10,11}$B + ${}^{209}$Bi~\cite{Gasques2009_PRC79-034605} are listed in 
Table~\ref{table:B}. For the reactions involving ${}^{10}$B, a mean value of 
$\bar\mu=-1.415$~fm$^{-1}$ is obtained, while for ${}^{11}$B, 
$\bar\mu=-1.900$~fm$^{-1}$. 
The mean value of $\mu=-1.415$~fm$^{-1}$ and $\mu=-1.900$~fm$^{-1}$ 
are used to calculate the CF cross sections for 
the reactions with ${}^{10}$B and ${}^{11}$B as projectiles, respectively. 
The comparison of the calculated CF cross sections to the experimental values 
are shown in Fig.~\ref{fig:B}. For these four reactions, the calculated 
CF cross sections shown by the pink dash-dotted line are in good agreement with 
the data. Moreover, $\sigma_{\rm Fus}$ scaled by the suppression factor 
$F_\mathrm{B.U.}=0.80$ for ${}^{10}$B and $F_\mathrm{B.U.}=0.91$ for ${}^{11}$B taken from 
Ref.~\cite{Wang2014_PRC90-034612} are represented by the blue dotted line, 
which coincide with the data. One can find that the suppression of CF cross 
sections for the reactions induced by ${}^{10}$B are independent of the target, 
as well as the reactions induced by ${}^{11}$B. Comparing the suppression of 
${}^{10}$B with that of its neighboring nucleus ${}^{11}$B, we find that the 
suppression factor for ${}^{11}$B is larger, as well as its breakup threshold.

\begin{table}[htb!] 
\caption{ %
Slope parameter $\mu$ obtained by making a least-squares fit to the
corresponding CF data using $\nu= 0.557-\mu R_0$ for reactions with 
${}^{10}$B and ${}^{11}$B as projectiles.
}
\begin{ruledtabular}
\begin{tabular}{ccccr}
& \multicolumn{2}{c}{${}^{10}$B}& \multicolumn{2}{c}{${}^{11}$B}\\
\cline{2-5}\noalign{\vskip3pt} 
  &$^{159}$Tb & ${}^{209}$Bi & $^{159}$Tb  &  ${}^{209}$Bi \\  
 \hline\noalign{\vskip3pt} 
$\mu$ (fm$^{-1}$) &$-$1.417 & $-$1.413 & $-$1.87 & $-$1.929 \\ 
\end{tabular} \label{table:B}
\end{ruledtabular}
\end{table}  

\subsection{\label{subsec:sys}Systematics of the prompt-breakup probability}

Based on the above analysis and discussions, one can find that both the 
logarithmic slope parameter $\mu$ of the prompt-breakup probability and the 
CF suppression for the reactions induced by the same nucleus are roughly 
independent of the target. In Ref.~\cite{Wang2014_PRC90-034612}, it was found 
that the suppression factor is mainly determined by the lowest 
breakup threshold energy of the projectile and an exponential relation between the 
suppression factor and the breakup threshold energy holds. 
Therefore, it is natural to explore the relation between the logarithmic slope 
$\mu$ and the breakup threshold.

\begin{figure}[htb!]
\centering{
\includegraphics[width=0.8\columnwidth]{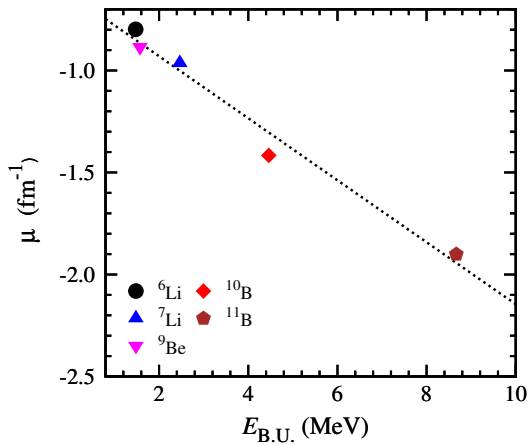}}
\caption{(Color online) The values of $\mu$ for ${}^{6,7}$Li, ${}^{9}$Be, 
and ${}^{10,11}$B as a function of the lowest breakup threshold energy $E_{\rm B.U.}$.
The dotted line denotes the empirical formula~(\ref{eq:emp1}).}\label{fig:aebu}
\end{figure}

The values of $\mu$ for ${}^{6,7}$Li, ${}^{9}$Be, and ${}^{10,11}$B as a 
function of theirs lowest breakup threshold energies $E_{\rm B.U.}$ are shown 
in Fig.~\ref{fig:aebu}. One can find that a linear relation between 
$\mu$ and $E_{\rm B.U.}$ is fulfilled, at least for 1.474 MeV $\le E_{\rm B.U.} 
\le$ 8.665 MeV. Furthermore, if the breakup threshold energy is large enough, 
the breakup effects would not affect the fusion and the absolute value of $\mu$ 
should be large enough to satisfy $P_{\rm B.U.}\approx 0$. 
An analytical formula that satisfies this physical limit is
\begin{equation}\label{eq:emp}
 \mu = -x-yE_{\rm B.U.},
\end{equation}
where $x$ and $y$ are parameters to be determined. By fitting the 
logarithmic slope parameter $\mu$ shown in Fig.~\ref{fig:aebu}, 
we get the values for these two parameters, $x=0.626~\text{fm}^{-1}$ and 
$y=0.152$  (MeV fm)$^{-1}$. That is, this analytical formula reads
\begin{equation}\label{eq:emp1}
 \mu = -0.626-0.152E_{\rm B.U.},
\end{equation}
where $E_{\rm B.U.}$ is in the unit of MeV. The logarithmic slope parameters 
obtained by this empirical formula are shown in Fig.~\ref{fig:aebu} by the 
dotted line. This analytical relation suggests that the effect of breakup on 
complete fusion may be indeed a threshold effect.
With Eq.~(\ref{eq:emp1}), the ECCBU model can be used to make predictions
of complete fusion cross sections for heavy ion reactions with
weakly bound nuclei as projectiles.

\section{\label{sec:summary}Summary}

The empirical coupled-channel model is extended by including the breakup effect 
which is described by a prompt-breakup probability 
function with two parameters, $\nu$ and $\mu$ [see Eq.~(\ref{eq:bu})]. The 
complete fusion suppression at above-barrier energies in reactions induced by 
the $^{9}$Be, $^{6,7}$Li and $^{10,11}$B projectiles on various targets are 
systematically investigated. For the 
reactions induced by $^{9}$Be, the parameters $\nu$ and $\mu$ have been extracted 
from the measured prompt-breakup probabilities, whereas for the other 
projectiles the parameter $\mu$ has been determined by making a fit to the complete 
fusion data. We found that both $\mu$ and the complete fusion suppression are 
roughly independent of the target for the reactions induced by the same 
projectile, $\mu$ being mainly determined by the lowest breakup threshold of 
the weakly bound projectile. 
An analytical formula which describes well the relation
between $\mu$ and the breakup threshold energy is proposed. 
It indicates that the effect of breakup on complete fusion is a threshold effect. 
Neglecting the sequential complete fusion (SCF) component, the present model suggests that 
the complete fusion suppression at above-barrier energies is roughly independent 
of the target in reactions involving the same weakly bound projectile, as this 
suppression is determined by the prompt breakup process.

\acknowledgements
We thank Arturo G\'omez Camacho for helpful discussions and for a careful reading 
of the manuscript.
This work has been partly supported by 
the National Key Basic Research Program of China (Grant No. 2013CB834400), 
the National Natural Science Foundation of China (Grants 
No. 11121403, 
No. 11175252, 
No. 11120101005, 
No. 11275248, and 
No. 11525524), and
the Knowledge Innovation Project of the Chinese Academy of Sciences (Grant No. 
KJCX2-EW-N01).
The computational results presented in this work have been obtained on 
the High-performance Computing Cluster of SKLTP/ITP-CAS and 
the ScGrid of the Supercomputing Center, Computer Network Information Center of 
the Chinese Academy of Sciences.


%

\end{document}